# Subset Queries in Relational Databases

Satyanarayana R Valluri
satya@iiit.net

Kamalakar Karlapalem
kamal@iiit.net

Centre for Data Engineering
International Institute of Information Technology
Gachibowli, Hyderabad, INDIA

## Abstract

In this paper, we motivated the need for relational database systems to support subset query processing. We defined new operators in relational algebra, and new constructs in SQL for expressing subset queries. We also illustrated the applicability of subset queries through different examples expressed using extended SQL statements and relational algebra expressions. Our aim is to show the utility of subset queries for next generation applications.

## 1 Introduction

Relational database systems currently support processing of tuples of relations to generate a single result as a set of tuples. Relational algebra, calculus and SQL are used to specify queries on relational databases. There have been extensions to relational algebra: 'group by' clause groups the tuples and each group is represented by a single aggregated result tuple, 'having' clause further restrics the groups that should be part of the result. A cube operator over $n$ dimensions generates $2^n$ result tuples, one for each group of tuples from a subset of dimension domain values. In all the operators: group by, having and cube, each group of tuples is represented by a single aggregated tuple. In this paper, we relax this notion by proposing subset queries and show the utility of subset queries in concisely specifying new class of user queries.

A subset is a set of tuples. A relation of subset is a set of subsets. The operations on a relation of subsets can be at two levels:

1. at the level of subsets treating them as atomic entities, and
2. at the level of tuples of subsets.

The objective of the paper is to present a class of queries that can be specified in a concise manner using relations of subsets. There are many applications that require such queries to be specified. It should be noted that these queries can be translated to standard SQL queries using complicated programming constructs which is non-trivial to application programmers. Therefore, there is a motivation to facilitate direct execution of subset queries.

Our aim in this paper is to reval the kind of applications that can be supported by subset queries but not to dwell on the efficiency issues of processing such queries. Once the notion of subset queries and their utility is accepted, additional work to efficiently execute subset queries can be done as it is a challenging open research problem.

In this paper, we

- motivated the need for relational database systems to support subset query processing,
- defined new operators in relational algebra, and new constructs in SQL for expressing subset queries, and
- illustrated the applicability of subset queries through different examples expressed using extended SQL statements and relational algebra expressions.

### 1.1 Motivation

We motivate the utility of subset queries through examples. Consider an Item relation in a grocery shop: (ItemId, Name, Weight, Price, Type) where ItemId is the unique id of the item, Name, Weight and Price denote the name, weight of the item in grams and the price of the item in Dollars respectively. The domain of Type is {"Eatable","Non-Eatable"}. Table 1 shows an extension of the Item relation.

**Query 1** *"What are the items whose weight is more than 50 and whose price is less than 50?".*

The answer to query 1 is the set of items $\{i_3, i_7, i_8\}$.

**Query 2** *"What are the sets of "Non-Eatable" items which can be bought such that the total price of these items is more than 150 and the total weight of these items is between 200 and 400?".*

Unlike the query 1, there are multiple results for query 2. The result of the query is the set of subsets: $\{\{i_2, i_9\}$, $\{i_1, i_2, i_9\}$,$\{i_1, i_4, i_9\}$, $\{i_4, i_7, i_9\}$, $\{i_1, i_4, i_7, i_9\}\}$.
Unfortunately, SQL does not have provision for specifying subset queries. This paper studies the problem of expressing subset queries and processing them. In the rest of the sections, we further develop the notion of subset queries and expound their utility.

The organization of the paper is as follows. Section 2 introduces relation of subsets and its properties. Section 3 develops subset relational algebra and extensions to SQL to specify subset queries. Section 4 presents recent literature on novel SQL extensions and contrasts it with subset queries. Finally, Section 5 presents conclusions.

# 2 Relation of Subsets

Let $R$ be the intension of a relational schema. Let $A = \{A_1, A_2, \ldots, A_m\}$ be the set of attributes of $R$. Let $r$=$\{t_1$,$t_2$,...,$t_n\}$ be an extension of $R$. Each $t_i, i = 1, 2, \ldots, n$ is a tuple defined over $R$. $t_i.A_j$ refers to the value of the attribute $A_j$ ($\in A$) of the tuple $t_i$ ($\in r$).

## 2.1 Terminology

In this section we formally define the subsets and the set of subsets called relation of subsets. Further, we develop the subset relational algebra.

**Definition 1** ***Subset of tuples:*** *A subset of tuples taken from extension r of relation R is denoted as* $s^r$. $|s^r|$ *is the cardinality of the subset, the number of tuples in the subset. Each tuple in* $s^r$ *is defined over the same intension R.*

The subset corresponding to query 1 is $s^r = \{i_3, i_7, i_8\}$

**Definition 2** ***Relation of subsets:*** *A relation of subsets* $\omega$ *is a set of subsets over the extension r of the relation R.* $\omega = \{s^r_1, s^r_2, \ldots, s^r_m\}$, *where each* $s^r_j$ *is a subset of tuples defined over the extension r,* $j = 1, 2, \ldots, m$, *where m is the number of subsets.*

For query 2, $\omega$ = { $\{i_2, i_9\}$, $\{i_1, i_2, i_9\}$,$\{i_1, i_4, i_9\}$, $\{i_4, i_7, i_9\}$, $\{i_1, i_4, i_7, i_9\}$ }.

## 2.2 Operations on subsets

In this section, we define the properties that can be defined over the subsets. These operations can be discussed under two categories: set operations and relational operations.

### 2.2.1 Set Operations

The set operations on subsets are shown in table 2.

### 2.2.2 Relational Operations

The relational operations on the subsets are similar to those of the standard relational operations on relations. The table 3 shows the various relational operations on subsets.

## 2.3 Properties of subsets

**Equality of subsets:** Two subsets $s^r_i$ and $s^r_j$ are said to be equal if they are defined on the same extension $r$ and contain the same set of tuples. $(s^r_i = s^r_j) \iff (\forall\, t_i \in s^r_i \iff t_i \in s^r_j)$.
**Complement of a subset:** Given $s^r$, the complement $\overline{s^r}$ is defined as the set of tuples in extension $r$ which do not belong to $s^r$. $\overline{s^r} = \{t_i \mid t_i \in r$ and $t_i \notin s^r\}$. $\overline{s^r} = r - s^r$.
Table 4 shows the properties of subsets based on the set theory.

## 2.4 Relation of subsets operations

The operations on the relation of subsets can be discussed under two categories: set operations and relational operations.

| Tuple | ItemId | Name | Weight | Price | Type |
|---|---|---|---|---|---|
| $i_1$ | 1 | Soap | 40 | 20 | Non-Eatable |
| $i_2$ | 2 | Face Powder | 250 | 70 | Non-Eatable |
| $i_3$ | 3 | Bread | 60 | 15 | Eatable |
| $i_4$ | 4 | Tooth Paste | 150 | 50 | Non-Eatable |
| $i_5$ | 5 | Jam | 35 | 65 | Eatable |
| $i_6$ | 6 | Chips | 25 | 18 | Eatable |
| $i_7$ | 7 | Hair Oil | 100 | 35 | Non-Eatable |
| $i_8$ | 8 | Sauce | 75 | 40 | Eatable |
| $i_9$ | 9 | Perfume | 60 | 100 | Non-Eatable |
| $i_{10}$ | 10 | Candy | 20 | 50 | Eatable |

Table 1: Extension of the Item relation

| Operation | Notation | Definition |
|---|---|---|
| **Union** $\cup$ | $s_i^r \cup s_j^r$ | $\{t_i \mid t_i \in s_i^r$ **or** $t_i \in s_j^r$ **or both**$\}$ |
| **Intersection** $\cap$ | $s_i^r \cap s_j^r$ | $\{t_i \mid t_i \in s_i^r$ **and** $t_i \in s_j^r\}$ |
| **Set Difference** $-$ | $s_i^r - s_j^r$ | $\{t_i \mid t_i \in s_i^r$ **but** $t_i \notin s_j^r\}$ |
| **Complement** | $\overline{s^r}$ | $\{t_i \mid t_i \in r$ **and** $t_i \notin s_r\}$ |

Table 2: Set Operations on **a subset of tuples**

#### 2.4.1 Set Operations

The table 5 displays the various set operations that can be applied on a relation of subsets. The output of the unary union and unary intersection is a single subset of tuples whereas the output of the rest of the operations is a relation of subsets. The name "cross" is appropriate for the cross union and cross intersection operations since every pair of subsets are considered similar to the cross product operation. While computing the cross union (cross intersection), if the result of the union (intersection) of two subsets is empty, then we do not include it in the output. But there may be some applications which require even the empty results in the output where a notion of outer cross union (cross intersection) can be used.

#### 2.4.2 Relational Operations

The table 6 shows the various relational operators that can be applied on relation of subsets. In [7], the notion of multi-relational algebra (MRA) is introduced for checking the correctness of query execution strategies in distributed databases. The cross cartesian product and the cross join operations are similar to the MJN operation defined on multi-relations.
As already mentioned in the previous sub-section the result of applying a select condition on a subset might be an empty set. We assume that such empty results are not included in the output. But we can define a notion of outer subset select which include such empty results also in the output. Similar extensions can be made to the cross join and the group by-having operations.

### 2.5 Relation of subsets properties

**Equality of relation of subsets:** Two relations of subsets $\omega_i$ and $\omega_j$ are said to be equal if they are defined on the same extension $r$ and both contain the same subsets.$\omega_i = \omega_j, \forall\ s^r \in \omega_i \iff s^r \in \omega_j$.
**Complement of relation of subsets:** Given a relation of subsets $\omega$, its complement, $\overline{\omega}$ contains the complements of the subsets of $\omega$. $\overline{\omega} = \{\overline{s^r} \mid \overline{s^r} = \overline{(s^r)} \forall s^r \in \omega\}$.

## 3 Supporting subset queries

### 3.1 Subset representation

A major problem for supporting subsets in relational databases is the representation of the subsets. Since the first normal form states that the domain of every attribute should take only atomic values [11], a subset cannot be represented as a single tuple.

| Operation | Notation | Definition | Description |
|---|---|---|---|
| Select | $\sigma_{C()}(s^r)$ | $\{t_i \mid t_i \in s^r$ and $t_i$ satisfies $C()\}$ C() is the select condition | Results in the tuples of $s^r$ that satisfy C() |
| Project | $\Pi_{<\text{attribute list}>}(s^r)$ | $\{t'_i \mid t'_i = \Pi_{<\text{attribute list}>}(t_i), \forall t_i \in s^r\}$ | Results in the tuples of the subsets whose columns are in projected attribute list |
| Cartesian Product | $s_a^{r_1} \times s_b^{r_2}$ | $\{t_1 \times t_2 \mid \forall t_1 \in s_a^{r_1}, \forall t_2 \in s_b^{r_2}\}$ | Results in the concatenation of every pair of tuples, $t_1 \in s_a^{r_1}, t_2 \in s_b^{r_2}$ |
| Join | $s_a^{r_1} \bowtie_{<jcl>} s_b^{r_2}$ | $\sigma_{<jcl>}(s_a^{r_1} \times s_b^{r_2})$ $\{t_1 \bowtie_{<jcl>} t_2 \mid \forall t_1 \in s_a^{r_1}, \forall t_2 \in s_b^{r_2}\}$ $< jcl >$ is the join condition list. | Results in the join of every pair of tuples, $t_1 \in s_a^{r_1}$,$t_2 \in s_b^{r_2}$. |
| Group By and Having | $\Im_{<gl>}(s^r)$ | $\Im_{<gl>}(t_1, t_2, \ldots, t_n)$,$\forall t_i \in s^r$ $< gl >$ is the group by attribute list | Results in the group by result applied on the tuples of $s^r$. Filters out the output with having clause. |

Table 3: Relational operations on **a subset of tuples**

| Property | Comment |
|---|---|
| $s^r \cup \overline{s^r} = r$ | |
| $s^r \cap \overline{s^r} = \phi$ | |
| $\overline{(\overline{s^r})} = s^r$ (assuming r contains only non-null tuples) | Double complement |
| $\overline{(s^{r_i} \cup s^{r_j})} = \overline{s^{r_i}} \cap \overline{s^{r_j}}$ $\overline{(s^{r_i} \cap s^{r_j})} = \overline{s^{r_i}} \cup \overline{s^{r_j}}$ | De Morgan's Laws |

Table 4: Properties of subsets

Hence, a subset is represented in the form of multiple tuples with a subset identifier (sid) to uniquely identify the members of the subset.

## 3.2 Subset query specification

The end user has to write multiple complex tedious standard SQL queries to specify subset queries. Hence the relational algebra and SQL needs to be extended to facilitate specification of subset queries.

**Relational algebra:** A new subset operator, $\omega$ which generates all subsets of tuples of a relation is introduced in relational algebra. The constraints to be satisfied by subsets are specified by a new subset select operator, $\sigma_{\subseteq}$.

**In SQL:** A clause "WITH SUBSETS" with an argument subset id is included in SQL with an optional "CONSTRAINED BY" clause that contains the constraints to be satisfied by the subsets.

**Example 1** *Table 7 shows the relational algebra expression and SQL for the query 2.*

## 3.3 Subset SQL

In this section, we introduce more extensions to relational algebra and SQL, and show how to write complex subset queries using them. The semantics of the operators are given in tables 5 and 6. For the sake of completeness, we begin the section with subset select operator.

| Operation | Notation | Definition | Explanation |
|---|---|---|---|
| Unary Union $\bigcup_U$ | $\bigcup_U(\omega)$ | $\{t_i \mid t_i \in \hat{s}^r, \hat{s}^r \in \omega\}$ | Results in a subset of tuples which appear in at least one of the subsets of $\omega$ |
| Unary Intersection $\bigcap_U$ | $\bigcap_U(\omega)$ | $\{t_i \mid \forall \hat{s}^r \in \omega, t_i \in \hat{s}^r\}$ | Results in a subset of tuples which appear in every subset of $\omega$ |
| Set Complement $\overline{\subseteq}$ | $\overline{\subseteq}(\omega)$ | $\{\overline{s^r} \mid s^r \in \omega\}$ | Results in the complements of the subsets of $\omega$ |
| Subset Union $\bigcup_\subseteq$ | $\omega_j \bigcup_\subseteq \omega_j$ | $\{s^r \mid$ where $s^r \in \omega_i$ or $s^r \in \omega_j\}$ | Results in the subsets which belong to $\omega_i$ or $\omega_j$ or both |
| Subset Intersection $\bigcap_\subseteq$ | $\omega_i \bigcap_\subseteq \omega_j$ | $\{s^r \mid$ where $s^r \in \omega_i$ and $s^r \in \omega_j\}$ | Results in the subsets which belong to both $\omega_i$ or $\omega_j$ |
| Cross Union $\bigcup_\times$ | $\omega_i \bigcup_\times \omega_j$ | $\{\hat{s}^r \mid \hat{s}^r = s_i^r \cup s_j^r$ where $s_i^r \in \omega_i, s_j^r \in \omega_j\}$ | Results in the union of every pair of subsets $s_i^r, s_j^r$ such that $s_i^r \in \omega_i$ and $s_j^r \in \omega_j$ |
| Cross Intersection $\bigcap_\times$ | $\omega_i \bigcap_\times \omega_j$ | $\{\hat{s}^r \mid \hat{s}^r = s_i^r \cap s_j^r$ where $s_i^r \in \omega_i, s_j^r \in \omega_j\}$ | Results in the intersection of every pair of subsets $s_i^r, s_j^r$ such that $s_i^r \in \omega_i$ and $s_j^r \in \omega_j$ |

Table 5: Set Operations on **a relation of subsets**

**Subset Select:** The unary subset operator $\sigma_\subseteq$ and the "CONSTRAINED BY" clause in SQL process the constraints to be satisfied by each subset in the set of subsets.

**Subset Project:** The unary subset project operation $\Pi_\subseteq$ projects a given set of attributes of each tuple in every subset. In SQL, the subset project attributes are specified in the "SELECT" clause.

**Example 2** *Consider a relation Shop: (ShopId, Location, Distance,Rating) whose extension is shown in table 8 where ShopId, Location and Distance are the id of the shop, location of the shop and the distance of the shop from a city center in Kilometers respectively. Rating is the popular rating of the shop given in a survey.*

*Query 1: "What are the locations of the shops visited by a customer if the total distance he travels is between 30 and 40 and if he visits only shops whose rating is more than 4.0?".*
*Query 2: "What is the total distance travelled and the maximum rating of the shop visited by a customer if the total distance he travels is between 30 and 40 and if he visits only shops whose rating is more than 4.0?".*
*The corresponding subset SQL queries, relational algebra expressions and results are shown in table 9.*

**Unary Union and Intersection:** The unary union operator $\bigcup_U$ (unary intersection operator $\bigcap_U$) gives the union(intersection) of all input subsets. In SQL, an additional clause, "APPLY UNARY UNION" ("APPLY UNARY INTERSECTION") is introduced which comes after the "WITH SUBSETS" and
"CONSTRAINED BY" clauses. It takes as input the subsets generated by the "WITH SUBSETS" clause or "CONSTRAINED BY" clause and outputs a single subset of tuples as the result.

**Example 3** *"What are the shops which are visited by* **ANY** *customer if the distance travelled by him is between 31 and 40 and if he visits only shops whose rating is more than 4.0?" and "What are the shops which are visited by* **EVERY** *customer if the distance travelled by him is between 31 and 40 and if he visits only shops whose rating is more than 4.0?". In the first query, the required answer is the union of the subsets which satisfy the given conditions where as in the second query it is the intersection of them. The corresponding subset SQL queries and the results are shown in table*

| Operation | Notation | Definition | Explanation |
|---|---|---|---|
| Subset Select $\sigma_{\subseteq}$ | $\sigma_{\subseteq <\text{scl}>}(\omega)$ <scl> is the select condition list | $\{\hat{s}^r \mid \hat{s}^r = \sigma_{<\text{scl}>}(s^r)$ $s^r \in \omega\}$ | Results in the subsets obtained by applying <scl> on each of the subsets of $\omega$ |
| Subset Project $\Pi_{\subseteq}$ | $\Pi_{\subseteq <\text{al}>}(\omega)$ <al> is the attribute list | $\Pi_{<\text{al}>}(s^r), \forall s^r \in \omega$ | Results in the subsets obtained by projecting <attribute list> on each of the subsets of $\omega$ |
| Cross Cartesian Product $\times_{\times}$ | $\omega_i \times_{\times} \omega_j$ | $s^{r_{ij}} \mid s^{r_{ij}} = s^{r_i} \times s^{r_j}$ $\forall s^{r_i} \in \omega_i, \forall s^{r_j} \in \omega_j$ | Results in the cross product of every pair of subsets, $s_i^r, s_j^r$ such that $s_i^r \in \omega_i$,$s_j^r \in \omega_j$ |
| Cross Join $\bowtie_{\times}$ | $\omega_i \bowtie_{\times <\text{jcl}>} \omega_j$ <jcl> is the join condition list | $s^{r_{ij}} \mid s^{r_{ij}} = s^{r_i} \bowtie_{<\text{jcl}>} s^{r_j}$ $\forall s^{r_i} \in \omega_i, \forall s^{r_j} \in \omega_j$ $\sigma_{\subseteq <\text{jcl}>}(\omega_i \times_{\times} \omega_j)$ | Results in the join of all pairs of subsets $s^{r_i}, s^{r_j}$ joined on <jcl> such that $s^{r_i} \in \omega_i$,$s^{r_j} \in \omega_j$ |
| Subset Group By $\Im_{\subseteq}$ | $\Im_{\subseteq <\text{gl}>}(\omega)$ <gl> is the group-by list | $\Im_{\subseteq <\text{gl}>}(s^r)$ $\forall s^r \in \omega$ | Results in the subsets of $\omega$ on which the group-by conditions are applied |

Table 6: Relational operations on **a relation of subsets**

| | |
|---|---|
| $\sigma_{\subseteq CA}(\sigma_{CN}(\omega(Item)))$ *CA: sum(Weight)>200∧sum(Weight)<400∧sum(Price)>150* *CN: Type="Non-Eatable"* | `SELECT * FROM Item WHERE Type="Non-Eatable"` **`WITH SUBSETS`** `Item sid` **`CONSTRAINED BY`** `sum(Weight)>200 and sum(Weight)<400 and sum(Price)>150` |

Table 7: Subset Queries

*10. The main subset query generates two subsets: $\{\{s_1, s_3\}, \{s_3, s_4\}\}$. Hence the result of the union is $\{s_1, s_3, s_4\}$ and that of the intersection is $\{s_3\}$.*

**Subset Union, Cross Union, Subset Intersection and Cross Intersection:** The subset union operator $\bigcup_{\subseteq}$ (subset intersection operator $\bigcap_{\subseteq}$) takes as input two sets of subsets and gives as output the union (intersection) of all the subsets in them. In SQL, "UNION" ("INTERSECTION") can be specified between two subset SQL queries. The cross union operator $\bigcup_{\times}$ (cross intersection operator $\bigcap_{\times}$) takes as input two sets of subsets and gives as output the union (intersection) of every pair of subsets, each one taken from each of the inputs. See table 5 for formal specification of these operators. A new SQL clause "CROSS UNION" ("CROSS INTERSECTION") is specified between two subset SQL queries.

**Example 4** *"What are the shops a customer might visit if the total distance travelled is between 30 and 36 and the rating of the shops visited is between 3.5 and 4.7* **or** *if the total rating of the shops visited is between 5.5 and 7.0 and if the distance of any shop to be visited is between 14 and 19?" and "What are the shops a customer might visit if the total distance travelled is between 30 and 36 and the rating of the shops visited is between 3.5 and 4.7* **and** *if the total rating of the shops visited is between 5.5 and 7.0 and if the distance of any shop to be visited is between 14 and 19?". The corresponding SQL queries for the above two queries is shown in table 11.*

*The results of the sub-queries (a) and (b) are $\{\{s_1, s_2\}, \{s_2, s_3\}\}$ and $\{\{s_2, s_5\}, \{s_3, s_5\}\}$ respectively. The result of the cross union query is $\{\{s_1, s_2, s_5\}$, $\{s_2, s_3, s_5\}$, $\{s_1, s_2, s_3, s_5\}\}$ and the result of the cross intersection operation is $\{\{s_2\}$,$\{s_3\}\}$. For the subset union query, the two queries are combined using "UNION" clause and the result of the query is* $\{\{s_1, s_2\}, \{s_2, s_3\}, \{s_2, s_5\}, \{s_3, s_5\}\}$.

| Tuple | ShopId | Location | Distance | Rating |
|---|---|---|---|---|
| s1 | 1 | M.G. Road | 20 | 4.5 |
| s2 | 2 | Airport | 15 | 3.9 |
| s3 | 3 | Downing Street | 18 | 4.6 |
| s4 | 4 | S.D. Road | 12 | 4.8 |
| s5 | 5 | Highway Road | 17 | 2.0 |

Table 8: Shop Table

<table>
<tr>
<td>
<code>SELECT sid,Location FROM Shop<br>WHERE Rating>4.0<br>WITH SUBSETS Shop sid<br>CONSTRAINED BY sum(Distance)>30 and<br>sum(Distance)<40</code>
<table>
<tr><th>sid</th><th>Location</th></tr>
<tr><td>1</td><td>M.G. Road</td></tr>
<tr><td>1</td><td>Downing Street</td></tr>
<tr><td>2</td><td>M.G. Road</td></tr>
<tr><td>2</td><td>S.D. Road</td></tr>
</table>
</td>
<td>
<code>SELECT sid,sum(Distance),max(Rating)<br>FROM Shop<br>WHERE Rating>4.0<br>WITH SUBSETS Shop sid<br>CONSTRAINED BY sum(Distance)>30 and<br>sum(Distance)<40</code>
<table>
<tr><th>sid</th><th>sum(Distance)</th><th>max(Rating)</th></tr>
<tr><td>1</td><td>38</td><td>4.6</td></tr>
<tr><td>2</td><td>32</td><td>4.8</td></tr>
</table>
</td>
</tr>
<tr><td>$\Pi_{\subseteq PA1}(\sigma_{\subseteq CA}(\sigma_{CN}(\omega(Shop))))$</td><td>$\Pi_{\subseteq PA2}(\sigma_{\subseteq CA}(\sigma_{CN}(\omega(shop))))$</td></tr>
<tr><td>PA1:sid,Location</td><td>PA2:sid,sum(Distance),max(Rating)</td></tr>
<tr><td colspan="2">CA: sum(Distance)>30∧sum(Distance)<40<br>CN: Rating>4.0</td></tr>
</table>

Table 9: Subset select and project queries and the results

**Cross Cartesian Product and Cross Join:** Given two sets of subsets R and S, the cross cartesian product $\times_\times$ generates a cartesian product of each pair of subsets taken from R and S, respectively. The cross join operator $\bowtie_\times$ is similar to the cross cartesian product except that each pair of subsets are joined on a given join condition. The cross cartesian product and cross join extend the standard relational cartesian product and join operations to sets of subsets, respectively. Example 5 and 6 show example cartesian and cross join queries.

**Example 5** *Consider the Item relation and the Shop relation example from example 2. Using the second query in example 4, consider the query: "Assuming that every item is available in every shop, give all the possibilities in which a customer visits a set of shops such that the distance of each shop is between 14 and 19 and the sum of the rating of the shops visited is between 5.5 and 7.0 and he buys the items whose price is less than 30 and the total weight of the items bought is between 60 and 90?". The SQL query for the above is shown in table 12.*
*The result of the first part of the query is* $\{\{s_2, s_5\},\{s_3, s_5\}\}$ *and the result of the second part of the query is* $\{\{i_1, i_6\},\{i_3, i_6\}\}$. *The cross cartesian product computes the cartesian product of every pair of subsets taking one subset from each. Hence there will be four subsets in the output.*

**Example 6** *Continuing the example 5, if there is a relation called Available with schema (ItemId, ShopId) which gives the information as to which item is available in which shop, then the query can be formulated as: " Give the possibilities in which a customer visits a set of shops such that the distance of each shop is between 14 and 19 and the sum of the rating of the shops visited is between 5.5 and 7.0 and he buys the items whose price is less than 30 and the total weight of the items bought is between 60 and 90?". A specific instance of the Available relation, the SQL query for the above statement and the result subsets are shown in table 13.*

**Subset Group By and Having $\Im_\subseteq$:** These operations are natural extensions to standard SQL group by and having expressions. In case of subsets, the group by and having clauses are applied to each

<table>
<tr>
<td>SELECT * FROM Shop<br>WHERE Rating>4.0<br>WITH SUBSETS Shop sid<br>CONSTRAINED BY<br>sum(Distance)>30 and<br>sum(Distance)<40<br>APPLY UNARY UNION<br>
<table>
<tr><th>sid</th><th>Tuple</th></tr>
<tr><td>1</td><td>$s_1$</td></tr>
<tr><td>1</td><td>$s_3$</td></tr>
<tr><td>1</td><td>$s_4$</td></tr>
</table>
</td>
<td>SELECT * FROM Shop<br>WHERE Rating>4.0<br>WITH SUBSETS Shop sid<br>CONSTRAINED BY<br>sum(Distance)>30 and<br>sum(Distance)<40<br>APPLY UNARY INTERSECTION<br>
<table>
<tr><th>sid</th><th>Tuple</th></tr>
<tr><td>1</td><td>$s_3$</td></tr>
</table>
</td>
</tr>
<tr>
<td colspan="2">$\bigcup_U(\sigma_{\subseteq CA}(\sigma_{CN}(\omega(Shop))))$ $\bigcap_U(\sigma_{\subseteq CA}(\sigma_{CN}(\omega(shop))))$<br>CA: sum(Distance)>30∧sum(Distance)<40<br>CN: Rating>4.0</td>
</tr>
</table>

Table 10: Subset queries using Unary Union and Unary Intersection

<table>
<tr>
<td>a)(SELECT * FROM Shop<br>WHERE Rating>3.5 and Rating<4.7<br>WITH SUBSETS Shop sid<br>CONSTRAINED BY sum(Distance)>30 and<br>sum(Distance)<36)<br>CROSS UNION<br>b)(SELECT * FROM Shop<br>WHERE Distance>14 and Distance<19<br>WITH SUBSETS Shop sid<br>CONSTRAINED BY sum(Rating)>5.5 and<br>sum(Rating)<7.0)</td>
<td>a)(SELECT * FROM Shop<br>WHERE Rating>3.5 and Rating<4.7<br>WITH SUBSETS Shop sid<br>CONSTRAINED BY sum(Distance)>30 and<br>sum(Distance)<36)<br>CROSS INTERSECTION<br>b)(SELECT * FROM Shop<br>WHERE Distance>14 and Distance<19<br>WITH SUBSETS Shop sid<br>CONSTRAINED BY sum(Rating)>5.5 and<br>sum(Rating)<7.0)</td>
</tr>
<tr>
<td>$R_1 \bigcup_\times R_2$</td>
<td>$R_1 \bigcap_\times R_2$</td>
</tr>
<tr>
<td>$R_1$: $\sigma_{CA1}(\sigma_{CN1}(\omega(Shop)))$<br>CA1: sum(Distance)>30∧sum(Distance)<36<br>CN1: Rating>3.5∧Rating<4.7</td>
<td>$R_2$: $\sigma_{CA2}(\sigma_{CN2}(\omega(Shop)))$<br>CA2: sum(Rating)>5.5∧sum(Rating)<7.0<br>CN2: Distance>14∧Distance<19</td>
</tr>
</table>

Table 11: Subset queries using Cross Union and Cross Intersection

subset in the set of subsets. Examples 7 and 8 show example group by and having queries.

**Example 7** *Consider the query on the items relation: "Find the set of items whose price is in the range of 40 and 70 and whose total weight is more than 500. In each such group, find the total price and minimum weight of the "Non-Eatable" and "Eatable" items". The corresponding SQL query and the result are shown in table 14.*

**Example 8** *In addition to the above query, if there is a condition which says, output only those sets whose total price is less than 110, then it is put in the HAVING clause. The SQL query and the result are shown in table 15.*

**Cardinality constraints:** The cardinality constraints on subset queries restrict the number of tuples in each of the subsets. The aggregate function clause **"count(sid)"** is used to specify such cardinality constraints. Example 9 illustrates cardinality constrained subset queries.

**Example 9** *Consider a query on the relation Item from example 1: "What are the set of "Eatable" items of cardinality 4 or 5 whose total weight is more than 190". The relational algebra expression and the SQL query for the above statement are shown in table 16. The results of the query are* $\{\{i_3, i_5, i_6, i_8\}$, $\{i_3, i_5, i_6, i_8, i_{10}\}\}$.

```
SELECT * FROM Item, Shop
WHERE Price<30
WITH SUBSETS Item sid,Shop sid
CONSTRAINED BY sum(Distance)>14 and
sum(Distance)<19
and sum(Rating)>5.5 and sum(Rating)<7.0
and sum(Weight)>60 and sum(Weight)<90
```

| Item | Shop | Item | Shop |
|---|---|---|---|
| $i_1$ | $s_2$ | ... | ... |
| $i_6$ | $s_2$ | $i_3$ | $s_3$ |
| $i_1$ | $s_5$ | $i_6$ | $s_3$ |
| $i_6$ | $s_5$ | $i_3$ | $s_5$ |
| ... | ... | $i_6$ | $s_5$ |

$(\sigma_{\subseteq CA1}(\omega(Shop)))\times_{\times}(\sigma_{\subseteq CA2}(\sigma_{CN2}(\omega(Item))))$

*CA1: sum(Distance)>14 ∧ sum(Distance)<19 ∧ sum(Rating)>5.5 ∧ sum(Rating)<7.0*

*CA2: sum(Weight)>60 ∧ sum(Weight)<90*

*CN2: Price<30*

Table 12: Subset query using cross cartesian product

**Available Relation**

| ItemId | ShopId |
|---|---|
| $i_2$ | $s_1$ |
| $i_3$ | $s_5$ |
| $i_6$ | $s_2$ |

```
SELECT * FROM Item, Shop, Available
WHERE Price<30
WITH SUBSETS Item sid,Shop sid
CONSTRAINED BY Item.ItemId = Available.ItemId
and Shop.ShopId = Available.ShopId
and sum(Distance)>14 and sum(Distance)<19
and sum(Rating)>5.5 and sum(Rating)<7.0
and sum(Weight)>60 and sum(Weight)<90
```

| Item | Shop |
|---|---|
| $i_6$ | $s_2$ |
| $i_3$ | $s_5$ |
| $i_3$ | $s_5$ |
| $i_6$ | $s_2$ |

$(\sigma_{\subseteq CA1}(\omega(Shop)))\bowtie_{\times JC1}(Available)\bowtie_{\times JC2}(\sigma_{\subseteq CA2}(\sigma_{CN2}(\omega(Item))))$

*JC1: Shop.ShopId = Available.ShopId*

*JC2: Item.ItemId = Available.ItemId*

*CA1: CA1: sum(Distance)>14 ∧ sum(Distance)<19 ∧ sum(Rating)>5.5 ∧ sum(Rating)<7.0*

*CA2: sum(Weight)>60 ∧ sum(Weight)<90*

*CN2: Price<30*

Table 13: The Available relation and the Subset Query using Cross Join

**Maximal and minimal subsets:** There will be some applications which require maximal or minimal subsets (in terms of cardinality) that satisfy a given set of constraints. For specifying maximal(minimal) subset constraint, the relational algebra operator $\hat{\omega}$ ( $\breve{\omega}$) is used. In SQL the WITH SUBSETS clause is followed by MAXIMAL (MINIMAL) to find the maximal (minimal) subsets. The MAXIMAL (MINIMAL) constraint is always accompanied by aggregate constraints on the subsets.

**Example 10** *Consider the following query on Item relation of example 1: "What are the maximal sets of "Eatable" items whose total weight is between 175 and 200?". The corresponding relational algebra expression and SQL statement are as shown in table 17. The results of the query are:* $\{\{i_3, i_5, i_6, i_8\}, \{i_3, i_5, i_8, i_{10}\}, \{i_3, i_6, i_8, i_{10}\}\}$. *Similar example can be constructed for minimal subsets.*

## 3.4 Summary of the SQL extension

The extended SQL syntax for select queries to support subset queries is given below where <table> <sid> pair denotes that the subset id for the subsets formed on <table> is <sid>. <cond> is a constraint to be imposed on the subsets.

# 4 Related Work

Customizing the relational databases in order to support more user preferences has received large interest from the researchers in the recent past. The effect of introducing powerset operator on the expressibility of the query languauges is widely studied in [22, 3, 31, 18, 19].Extending SQL with more clauses in order to enable the end user to write more

```
SELECT sid,Type,sum(Price),min(Weight)
FROM Item WHERE Price≥40 and Price≤70 WITH
SUBSETS Item sid CONSTRAINED BY
sum(Weight)>500 GROUP BY Type
```

| *sid* | *Type* | *sum(Price)* | *min(Weight)* |
|---|---|---|---|
| *1* | *Non-Eatable* | *120* | *150* |
| *1* | *Eatable* | *105* | *35* |
| *2* | *Non-Eatable* | *120* | *150* |
| *2* | *Eatable* | *150* | *20* |

$$\Pi_{\subseteq PA}(\Im_{\subseteq Type}(\sigma_{\subseteq CA}(\sigma_{CN}(\omega(Item)))))$$

*PA: sid,Type,sum(Price),min(Weight)*

*CA: sum(Weight)>500*

*CN: Price≥40 ∧ Price≤70*

Table 14: Subset Query using Group By

```
SELECT sid,Type,sum(Price),min(Weight)
FROM Item WHERE Price≥40 and Price≤70 WITH
SUBSETS Item sid CONSTRAINED BY
sum(Weight)>500 GROUP BY Type Having
sum(Price)<110
```

| *sid* | *Type* | *sum(Price)* | *min(Weight)* |
|---|---|---|---|
| *1* | *Eatable* | *105* | *35* |

$$\Pi_{\subseteq PA}(\Im_{\subseteq Type,HC}(\sigma_{\subseteq CA}(\sigma_{CN}(\omega(Item)))))$$

*PA: sid,Type,sum(Price),min(Weight)*

*HC: sum(Price)<110*

*CA: sum(Weight)>500*

*CN: Price≥40 ∧ Price≤70*

Table 15: Subset Query using Group By and Having

```
SELECT ... FROM ... WHERE ...
WITH SUBSETS <table> <sid>,...,
                       <table> <sid>
[CONSTRAINED BY <cond>,...,<cond>]
[[UNARY UNION]|
            [UNARY INTERSECTION]]
[[GROUP BY ...] [HAVING ...]]

SQ(CROSS UNION|
CROSS INTERSECTION)SQ
(where SQ is a subset SQL query).
```

personalized queries [6, 4, 39, 30, 36, 33, 10, 16], developing the relational algebra and SQL for user preferences [32, 17, 2, 21, 35, 29, 26, 27, 9, 8, 42, 41, 20, 28, 43], processing queries on set-oriented attributes in object relational databases [38, 37], finding top-K results when multiple preference functions are provided [12, 14, 13, 15, 5], joining multiple ranked inputs [34, 23] and finding top-K results of joins on multiple inputs [24] are some of the efforts the authors are aware of in this direction. Inspite of all these efforts, the subset query problem as elaborated in this paper is yet to be addressed to the best of our knowledge.

### 4.1 Powerset Operator

Some of the disadvantages of relational model are lack of semantics and the fact that it forces the data to have a flat structure that the real data does not always have. [22], [3] and [31] proposed generalized data models. [22] proposes a ”format model” that generalizes the relational and hierarchical models. They model database schemes as trees, where each leaf represents data and each internal node represents some connection between the data. But they ignored the issue of a data manipulation language. [3] describes ”database logic”, a mathematical model for databases that generalizes relational, hierarchical and network models but the query language proposed enables one to write even non-computable queries. [31] proposes a data model that generalizes all three principal models. It models the database scheme as a directed graph, where leaves represent data and internal nodes represent connections between the data. They proposed a non-procedural query language and an algebraic query language for the data model and showed that they are equivalent.

[31] introduced the *powerset operator* in the algebraic language proposed. [1] presents a gen-

| | |
|---|---|
| $\sigma_{\subseteq CA}(\sigma_{CN}(\omega(Item)))$<br>*CA: count(sid)≥4∧count(sid)≤5∧sum(Weight)>190*<br>*CN: Type="Eatable"* | `SELECT * FROM Item`<br>`WHERE Type="Eatable"`<br>`WITH SUBSETS Item sid`<br>`CONSTRAINED BY sum(Weight)>190 and`<br>`count(sid)>=4 and count(sid)<=5` |

Table 16: Subset query using cardinality constraint

| | |
|---|---|
| $\sigma_{\subseteq CA}(\hat{\omega}(\sigma_{CN}(Item)))$<br>*CA: sum(Weight)>175∧sum(Weight)<200*<br>*CA: Type="Eatable"* | `SELECT * FROM Item WHERE Type="Eatable"`<br>`WITH SUBSETS Item sid MAXIMAL`<br>`CONSTRAINED BY sum(Weight)>175 and`<br>`sum(Weight)<200` |

Table 17: Subset query using maximal set constraint

eral model for complex values and languages for it. The main result of the [1] is that the domain-independent calculus, the safe calculus, the algebra and the logic-programming oriented language have the equivalent expressive power. To make the algebra proposed in [1] equivalent to the calculus, the *powerset operator* has to be included in it. But including the powerset operator allows one to express queries that cannot be computed in PTIME in the size of the database. Hence, they propose a notion of safety and they prove that the restricted calcus corresponds to the algebra without the powerset operator.

In [18], the augmentation of the nested relational algebra with programming constructs such as while-loops and for-loops is discussed. They show that the extensions yield a query language equivalent to the powerset algebra. In [19], it has been shown that the least fixpoint clusure of the nested algebra is equivalent to the powerset algebra.

The notion of subset introduced in the paper is completely different from the nested relational model. In all this work, the aspects pertaining to the expressibility and the power of the query languages is studies. None of the work studied modelling representation of subset of tuples in the context of a relational database and the algorithms for processing queries on them.

## 4.2 Extending SQL

[6] extended SQL with an optional STOP AFTER clause in order to facilitate limiting the cardinality of a query result. In order to integrate STOP AFTER query processing seamlessly into an existing query processor, the paper proposes a new query operator called *Stop*. Two implementations of the *Stop* operator are proposed, Scan-Stop and Sort-Stop. Two different approaches, Conservative and Aggressive, are proposed to generate the query plans for STOP AFTER queries.

[4] introduced a new operator called *Skyline* operator. When multiple conflicting preferences are present, the Skyline operator finds the tuples which are not comparable based on the preferences. A notion of domination is developed in which a tuple dominates another tuple if it is as good or better in all dimensions and better in at least one dimension. The result of the skyline query is set of points which are not dominated by any other point. An extension to SQL is proposed in the form of an optional SKYLINE OF clause. In [4], block-nested-loop algorithm and a divide-and-conquer algorithm are proposed for processing skyline queries along with other miscellaneous algorithms like using B-tree, R-tree are.

In [39], efficient algorithms for progressive computation of skyline queries are studied. Two algorithms, Bitmap-based and B+-tree based are proposed. The bitmap-based algorithm is non-blocking and exploits a bitmap structure to quickly identify whether a point is interesting or not. The B+-tree based method uses a transformation mechanism and B+-tree index to return the skyline points in batches. [30] proposed a nearest neighbor based algorithm called NN algorithm. The main features of this algorithm are: it is an on-line algorithm, and the nearest neighbor search is a well-studied operation and hence efficient algorithms are available for it. The NN algorithm repeatedly partitions the search space based on nearest neighbors and quickly prunes the non-skyline points. A branch-and-bound algorithm for skyline computation is proposed in [36]. It scales better than the NN algorithm in terms of performance and space requirements. It uses an R-tree and a heap to prune the non-skyline points by traversing the nodes of the R-tree and ordering them.

[33] proposed a new divide-conquer algorithm, DCSkyline for 2D skyline queries which is more scalable when compared to BBS and NN algorithms. It reduces the number of paths the BBS algorithm has to examine in the R-tree in order to prune the non-skyline points. Also, DCSkyline algorithm eliminates the need for a global validation every time a candidate skyline point is generated.

[16] proposed Sort-Filter-Skyline algorithm where in the tuples are sorted using an external sorting algorithm and a window of tuples are compared with the sorted tuples to prune the non-skyline tuples. This algorithm is quite similar to the block-nested-loop algorithm of [4].

The problem of cardinality estimation for skyline queries is studied in [16]. Under a basic set of assumptions on the attributes and their domain values, the paper calculates the expected skyline cardinality based on the harmonics of the number of tuples and/or the number of dimensions of the data.

## 4.3 Preferences in SQL

The earliest work we are aware of on adding preferences to databases is [32]. A new clause "prefer" is added to the query language which enables the user to specify simple and complex preferences. [17] discusses the use of preferences in Datalog. The paper describes a bottom-up evaluation method for preference datalog programs. The notion of preference functions is formally defined in [2]. The paper presents a framework for expressing and combining multiple such preference functions. [21] extended this work and provided algorithms to implement the framework. It materializes in advance multiple views in order to provide short response time to client queries.

In [35], the relational data model is extended to incorporate partial orderings. Partially ordered relational algebra is developed and proposes Ordered SQL (OSQL) as a query language for ordered databases. In [8, 9], a logical framework for formulating *intrinsic* preferences, the preferences which use only in-built predicates, is proposed. A new relational operator called *winnow* is introduced which selects from its argument relation the most preferred tuples according to the given preference relation.

[26] models the preferences as strict partial orders. Various preference constructors are defined to specify complex preferences. A preference algebra is also developed over these constructs. To process the preference queries, Best-Matches-Only (BMO) query model is proposed which decomposes complex preference queries into simpler ones, and uses divide and conquer algorithms. Preference SQL, an extended SQL for expressing the preferences using the strict partial order model is proposed in [29]. Several new clauses like "PREFERRING" and "BUT ONLY" are added to the standard SQL. Preference SQL is implemented as an intermediate layer between the application and the SQL database system. Algebraic optimization of preference SQL queries is discussed in [27, 28]. Several transformation rules are proposed which generate better query execution plans. Mining of the preferences in user log data is discussed in [20].

Similar to the *winnow* operator of [8, 9], [42] uses *Best* operator to express preferences. A preference graph is constructed based on the preference function given by the user. In [41], algorithms for implementing the Best operator are discussed. In addition to this, a special data structure called $\beta$-tree is presented which speeds up the computation of Best operator. Efficient maintenance of $\beta$-trees in the presence of database updates is discussed in [43].

The problem of set-containment joins is studied in [38]. A partitioning based algorithm is proposed which uses a replicating multi-level partitioning scheme based on a combination of set elements and signatures.

## 4.4 Top-K queries

In multimedia applications, sometimes the results of a query are "fuzzy" instead of "exact". The results of the query are ranked based on a ranking function. The basic Fagin algorithm for combining such fuzzy information from multiple sources and obtaining the top-K results is discussed in [12]. A more elegant algorithm called "Threshold Algorithm" is proposed in [14, 15]. An approximate algorithm based on the threshold algorithm is pro-posed in [13]. [5] discusses the processing of top-k selection queries. Given a query to find the top-K closest tuples to a given tuple based on a distance function, all the points which fall within the distance of a radius are retrieved and the result of the query is computed. The radius used to retrieve the tuples is calculated by exploiting the statistics of the data.

### 4.5 Joining ranked inputs and Top-K join results

Incremental join of multiple ranked inputs is studied in [34]. J* algorithm proposed in the paper allows user-defined join predicates, multiple join levels and nested join hierarchies. Approximate versions of the algorithms is also proposed that trades output accuracy for reduced database access costs and space requirements. [23] introduces a new pipelined query operator NRA-RJ, that produces global rank from input ranked streams based on a score function. Processing of top-K join queries is studied in [24]. A variant of ripple join algorithm is used to implement two new physical query operators for the rank-join algorithm.

### 4.6 Discussion

In contrast to above related work, the subset queries are quite different from set valued attributes and nested relational algebra related work.

- The idea of set-valued attributes [38, 37] is also different from that of subsets. Each attribute can take as values a set of values where as the subset allows us to refer to a subset of tuples as a single entity.
- The notion of subsets introduced in this paper is completely different from that of the nested relational model [25, 40]. In nested relational model, the value of an attribute can be a set of values or a hierarchy of value or a relation by itself. On the other hand, in this paper, the standard relational model is used while modelling the notion of referring or accessing a relation of subsets of tuples.

## 5 Conclusions

There are many applications that require multiple subsets of a relation to be generated and processed. Current database management systems do not have this capability. The closest such functionality available for users is 'group by' clause, and data cube support. The notion of processing subsets of tuples of a relation is challenging. In this paper, we presented subset relational algebra and extended SQL to express subset queries with the aim towards bringing out the power of subset queries to express wide range of novel application queries. The issues of processing subset queries is a grand challenge and we are currently working on efficiently executing maximal subset queries.